\documentclass[journal,12pt,onecolumn,draftclsnofoot]{IEEEtran}
\usepackage{graphicx}
\usepackage{epsf}
\usepackage{mathtools}
\usepackage{amssymb}
\usepackage{amsmath, amsfonts}
\usepackage{latexsym}
\usepackage{subfigure}
\usepackage{multirow}
\usepackage{multicol}
\usepackage[table]{xcolor}
\usepackage{color}

\usepackage{tabularx}
\usepackage{lipsum}

\usepackage{algorithm}
\usepackage{algorithmic}

\usepackage{mathrsfs}

\definecolor{orange}{RGB}{255,127,0}
\definecolor{blue}{RGB}{0,0,255}
\definecolor{red}{RGB}{220,0,0}
\definecolor{green}{RGB}{0,120,0}
\definecolor{grey}{RGB}{255,120,255}

\begin{document}
{
\title{\LARGE{Coexistence of Outdoor Wi-Fi and Radar at 3.5 GHz}}

\author
{
Seungmo Kim and Carl Dietrich

\thanks{S. Kim and C. Dietrich are with the Bradley Department of Electrical and Computer Engineering, Virginia Tech in Blacksburg, VA, USA (e-mail: \{seungmo, cdietric\}@vt.edu).}
}

\maketitle

\begin{abstract}
Coexistence between radar and outdoor Wireless Fidelity (Wi-Fi) needs thorough study since the IEEE 802.11 Working Group (WG) opposed the latest rules in 3550-3700 MHz (the 3.5 GHz band) that require ``exclusion zones.'' This letter proposes a method that suppresses Wi-Fi-to-radar (WtR) interference, in which a Wi-Fi transmitter (TX) is selected to avoid beam angles toward the victim radar. It is distinguished from prior schemes since it ensures that the Wi-Fi remains operable while suppressing the WtR interference.
\end{abstract}


\IEEEpeerreviewmaketitle

\section{Introduction}
Since the United States Federal Communications Commission (FCC) released new rules for shared use of the 3.5 GHz band \cite{ro_35g}, IEEE 802.11 WG has been objecting to the decision that exclusion zones must exist for protecting federal radars, since Wi-Fi cannot bring enough benefit without serving users living in the coastal areas where a large population of this nation resides \cite{fasttrack}. This communication-radar coexistence was discussed in some recent literature \cite{35g_model_zander_lett11}-\cite{uwashington16}. More recent work proposed interference reduction techniques \cite{zander_dyspan14}-\cite{hanbat16}.

While little work exists beyond the prior work above, this letter proposes a WtR interference mitigation technique, assuming that the Wi-Fi is deployed \textit{outdoors} and thus adopts \textit{directional antennas}. The growing demand for affordable mobile broadband connectivity is driving the development of Heterogeneous Networks (HetNets) where macro cells will be complemented by a multitude of small cells, which will require broader deployment of outdoor Wi-Fi \cite{outdoor_nokia16}. The outdoor deployment is expected to be the major source of WtR interference because it uses higher transmit power and antenna gain due to directional antennas for connecting cells and networks.

The contributions of this letter are as follows:
\begin{itemize}
\item{It characterizes coexistence of radar with Wi-Fi, rather than the Long-Term Evolution (LTE) as in \cite{hindawi16}-\cite{hanbat16}. This letter will serve a critical need in the 3.5 GHz band coexistence: \textit{reduction of the exclusion zone}. Moreover, Wi-Fi is more complicated to model because the TX is randomly selected between an access point (AP) and a normal station (STA), whereas transmission of base stations and user equipments are strictly divided in LTE.}
\item{It proposes a method that mitigates WtR interference. Our work is distinguished from \cite{zander_elsevier13}-\cite{hanbat16} since it (i) requires no Wi-Fi network to stop transmission during an interference suppression period and (ii) maintains acceptable Wi-Fi performance while suppressing WtR interference.}
\item{It proposes a comprehensive protocol that enables to (i) acquire the location and (ii) piggyback the location report on the channel sounding, based on the up-to-date 3.5 GHz rules. This distinguishes this letter from \cite{zander_elsevier13} that unrealistically assumed perfect synchronization between the radar and the Wi-Fi.}
\end{itemize}

\section{Analysis of Interference}
In this section, we provide an analysis framework for WtR and radar-to-Wi-Fi (RtW) interference.

\begin{table}[t]
\caption{Summary of Key Notation}
\centering
\begin{tabular}{c l}
\hline
\textbf{Notation} & \textbf{Description}\\
\hline
$\mathtt{x} = \left(x, y\right)$ & Position of a node\\
$\lambda$ & Density (the number of points) of a PPP\\
$\theta_w, \theta_r$ & \textit{Off-axis angles} of a Wi-Fi node and the radar, respectively (See Fig. \ref{fig_geometry_ppp})\\
$\Theta$ & Threshold that limits a $\theta_w$ in the proposed method\\
$\mathcal{S}_{\theta_w}, \mathcal{S}_p$ & Sets of Wi-Fi nodes sorted in $\theta_w$ and the priority, respectively\\
\hline
\end{tabular}
\label{table_notation}
\end{table}

\subsection{System Model}\label{sec_system_model}
\subsubsection{Geometry}\label{sec_system_model_geometry}
One radar is placed at the origin of the quadrant, $\mathcal{O} = \left(0,0\right)$. The radar beam rotates with a revolution rate of $\rho$ rotations per minute (rpm). The distance between the radar and the center of a ``Wi-Fi region'' is denoted by $d$. A \textit{Wi-Fi network} is composed of one ``fixed'' AP with multiple STAs attached to the AP. A \textit{Wi-Fi region} refers to a region with multiple Wi-Fi networks. In a geometry given in Fig. \ref{fig_geometry_ppp}, a Wi-Fi region is denoted by $\mathbf{R}^2_{reg}$ and a Wi-Fi network is denoted by $\mathbf{R}^2_{net}$. Distribution of $\lambda_{ap}\left(>0\right)$ Wi-Fi APs in a $\mathbf{R}^2_{reg}$ follows a Poisson Point Process (PPP): $\mathtt{x}_{ap}=\left(x_{ap},y_{ap}\right) \in \mathbf{R}^2_{reg}$. Then, a Wi-Fi network is formed around each AP, which is expressed as $\left|\mathbf{R}^2_{net,k}\right| = \left|\mathtt{x}_{ap}-\mathtt{x}_{sta}\right| \le r_{net}$ where $\mathbf{R}^2_{net,k}$ is the $k$th Wi-Fi region in a $\mathbf{R}^2_{reg}$ and $r_{net}$ is the radius of a $\mathbf{R}^2_{net,k}$ and it is kept constant among different $\mathbf{R}^2_{net,k}$'s. Note that an $\mathbf{R}^2_{net,k}$ is circular although an AP antenna uses beamforming, since it represents a geometry that an AP can serve. Distribution of $\lambda_{sta}\left(>0\right)$ STAs is represented as another PPP: $\mathtt{x}_{sta}=\left(x_{sta},y_{sta}\right) \in \mathbf{R}^2_{net}$. As such $\lambda_{sta}$ and $\lambda_{ap}$ are regarded as the \textit{densities} of the PPPs that are defined in $\mathbf{R}^2_{net,k}$ and $\mathbf{R}^2_{reg}$, respectively. Note that $\mathtt{x}_{sta}$ and $\mathtt{x}_{ap}$ are homogeneous point processes where $\lambda_{sta}$ and $\lambda_{ap}$ are constant in different $\mathbf{R}^2_{net,k}$'s and $\mathbf{R}^2_{reg}$'s, respectively. Therefore $\mathtt{x}_{sta}$ and $\mathtt{x}_{ap}$ are distributed uniformly on X and Y axes on a quadrant \cite{daley}.

We assume that antennas of the Wi-Fi TX and receiver (RX), and the radar are at the same height, which excludes the elevation plane from consideration. This assumption is reasonable because coexistence likely occurs along the coast, where the Wi-Fi networks are deployed at almost the same height from the sea level. Now, on the azimuth plane, an \textit{interference axis} is defined as the line connecting an interferer TX and a victim RX. Since both the radar and the Wi-Fi use directional antennas, an interference level is dominantly determined by an angle of a beam relative to an interference axis, namely an \textit{off-axis angle}. As in Fig. \ref{fig_geometry_ppp}, $\theta_{w}$ and $\theta_{r}$ denote off-axis angles of a Wi-Fi TX and the radar, respectively.

\subsubsection{Antenna beam patterns}
The antenna gain for the radar is based on a \textit{high-gain antenna model} \cite{ntia} with 22 $< G_{max} = 33.5 <$ 48 dBi where $G_{max}$ is a maximum antenna gain and the value of 33.5 dBi comes from a relevant benchmark \cite{uwashington16}.

For the Wi-Fi's radiation pattern, we adopt the general \textit{linear array} which is given by \cite{antenna}
\begin{align}\label{eq_beam_pattern_wifi}
G\left(\theta\right) = G_{max} - \exp\left( -2\pi\j \delta \sin\theta \right)
\end{align}
where $\delta$ denotes the antenna element separation distance that is half a wavelength, and $\theta$ denotes an azimuth angle. A Wi-Fi antenna (for both the AP and the STA) is composed of 4 elements that are placed horizontally linearly. The maximum antenna gain for an element is 2.15 dBi, which results in $G_{max} = 2.15 + 10\log_{10}4 \approx 8.17 \text{ dBi}$.

\begin{figure}[t]
\centering
\includegraphics[width = 0.8\linewidth]{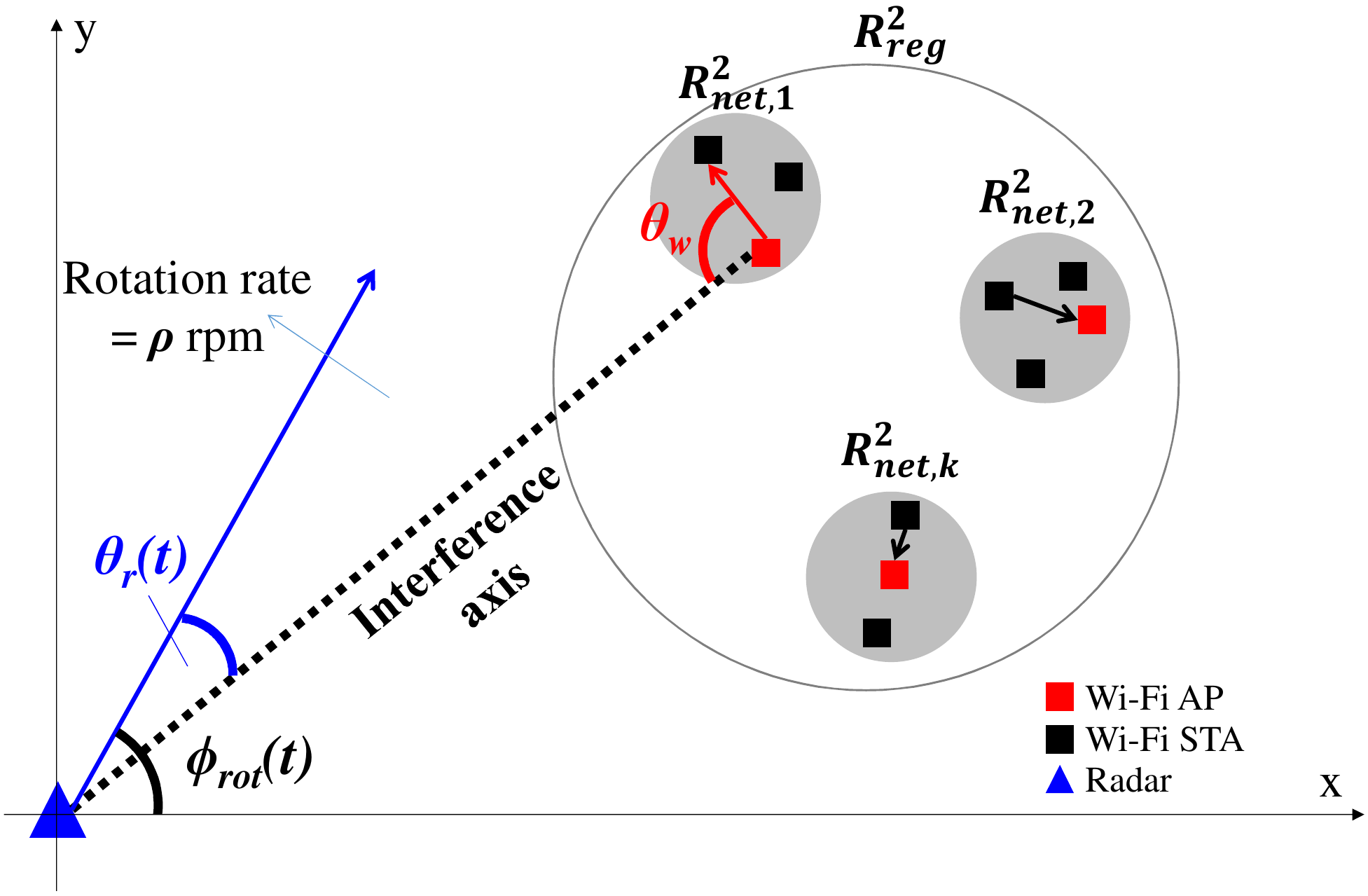}
\caption{Geometry of a radar-Wi-Fi coexistence}
\label{fig_geometry_ppp}
\end{figure}

\subsubsection{Multiple access in Wi-Fi}
In a Wi-Fi network, the AP and STAs compete for the medium in two different manners, enhanced distributed channel access (EDCA) and normal carrier-sense multiple access (CSMA), respectively. This letter approximates a scheme where a priority value ranges $0 \le \mathtt{p} \le 7$ as in EDCA. Reflecting recent practical Wi-Fi environments where an AP requires more chance of transmission, the priorities are ranged in $4 \le \mathtt{p} \le 7$ for an AP and in $0 \le \mathtt{p} \le 7$ for a STA. The priority values are uniformly randomly distributed within a range. Note that although there is prior work thoroughly characterizing EDCA, the focus of this letter is the proposition of a new technique for mitigating WtR interference. As such it proposes a metric that reasonably approximates the performance of a Wi-Fi network, reflecting the difference between EDCA and normal CSMA. The metric is called \textit{normalized priority-based performance indicator ($\mathtt{NPPI}$)} which is given by
\begin{align}\label{eq_nppi}
\mathtt{NPPI} = \begin{cases}\mathtt{SINR} * \left(\mathtt{p}+1\right)/8, &\text{EDCA}\\
\mathtt{SINR}, &\text{Normal CSMA}
\end{cases}
\end{align}
where signal-to-interference-plus-noise ratio ($\mathtt{SINR}$) at a Wi-Fi RX shall be discussed in (\ref{eq_sinr_no_mitigation}). In the normal CSMA, the AP and the STAs within a network has equal chance of transmission.

\subsection{Interference calculation}\label{sec_system_model_interference}
\subsubsection{Wi-Fi-to-radar (WtR) interference}\label{sec_system_model_interference_wtr}
Based on the geometry shown in Fig. \ref{fig_geometry_ppp}, we can formulate an interference power that is received at a radar location in $\mathbf{R}^2_{reg}$ from an ``individual'' Wi-Fi TX at a time instant, $t$, which is given by
\begin{equation}\label{eq_i_individual}
I\left(\mathtt{x}_{ap},\mathtt{x}_{sta},t\right) = l\left(\|\overrightarrow{\mathcal{OQ}}\|\right) P_T G_T\left(\theta_w\right) G_R\left(\theta_r\left(t\right)\right)
\end{equation}
where $G_T$ and $G_R$ denote TX and Rx antenna gains corresponding a Wi-Fi TX and the radar in this case, respectively. Also, $l\left(\cdot\right) = 259\|\overrightarrow{\mathcal{O}Q}\|^{-3.97}$ is a path loss \cite{uwashington16} between the origin of the quadrant $\mathcal{O}$ (location of the radar) and a point $\mathcal{Q} = \mathtt{x}_{ap} \text{ or } \mathtt{x}_{sta}$. Note that (\ref{eq_i_individual}) is a function of $\mathtt{x}_{ap}$, $\mathtt{x}_{sta}$, and $t$ because they determine $\theta_w$ and $\theta_r$. With an AP as an example as depicted in Fig. \ref{fig_geometry_ppp}, $\theta_{w} = \cos^{-1} \frac{\overrightarrow{\mathtt{x}_{ap}\mathtt{x}_{sta}} \cdot \overrightarrow{\mathtt{x}_{ap}\mathcal{O}}}{\|\overrightarrow{\mathtt{x}_{ap}\mathtt{x}_{sta}}\|\|\overrightarrow{\mathtt{x}_{ap}\mathcal{O}}\|}$ and $\theta_{r}\left(t\right) = \cos^{-1} \frac{\overrightarrow{\mathcal{O}\mathtt{x}_{ap}} \cdot \overrightarrow{\mathcal{O}\mathcal{O}'}}{\|\overrightarrow{\mathcal{O}\mathtt{x}_{ap}}\|\|\overrightarrow{\mathcal{O}\mathcal{O}'}\|}$ where $\mathcal{O}'$ is a reference point to indicate the radar beam's direction. Note that it is given by $\mathcal{O}' = \left(d\cos\phi_{rot}\left(t\right), d\sin\phi_{rot}\left(t\right)\right)$. A radar beam rotation angle, $\phi_{rot}\left(t\right)$, is a function of time and is given by $\phi_{rot} \left(t\right) = \frac{2\pi\rho}{60}t$ where $\rho$ is recalled to be a revolution rate (the number of rotations per minute) of a radar, and $t$ is a time instant measured in seconds. Note that $P_T$ is differentiated according to whether the Wi-Fi TX being an AP or a STA. The probability that an AP or a STA becomes TX depends on the multiple access schemes, EDCA or normal CSMA.

It is very important to note that a WtR interference is composed of an \textit{aggregate} signal power received by multiple Wi-Fi TXs simultaneously. From (\ref{eq_i_individual}), an aggregate interference that is received by a radar located at $\mathcal{O}$ from all of the Wi-Fi networks in $\mathbf{R}^2_{reg}$ at a time instant $t$ can be formulated as
\begin{align}\label{eq_i_aggregate}
I_{wtr}\left(\mathtt{x}_{ap},\mathtt{x}_{sta},t\right) = P_T \displaystyle \sum_{\mathtt{x}_{ap} \in \mathbf{R}^2_{reg}} l\left(\|\overrightarrow{\mathcal{O}Q}\|\right) G_T\left(\theta_w\right) G_R\left(\theta_r\left(t\right)\right).
\end{align}

\noindent For a PPP of density $\lambda$, Campbell's theorem \cite{stoyan} offers a way to calculate a mean of a sum of an arbitrary real-valued function $h\left(\cdot\right)$ over a point process $\mathcal{S}$ on a $d$-dimensional region $\mathbf{R}^d$ is given by $\mathbb{E}\left[\sum_{u\in\mathcal{S}}{h\left(u\right)}\right] = \lambda \int_{\mathbf{R}^d} h\left(u\right) du$.

Note that a radar's operation must be completely protected since it serves the national security, which requires theoretically ``zero'' possibility of violation of the RtW interference threshold (set to -10 dB). Hence, we identify the \textit{maximum interference power} during a radar's rotation that is averaged over all possible $\mathtt{x}_{ap}$ and $\mathtt{x}_{sta}$. The formal definition of a radar rotation time is $t_{n} \le t \le t_{n+1}$ where $t_{n}$ is the time at which the $n$th rotation is completed and thus $t_{n+1} - t_{n}$ represents a rotation time. This leads to a \textit{mean maximum aggregate interference ($\mathtt{MMAI}$)} as
\begin{align}\label{eq_i_aggregate_mean}
&\mathbb{E}\left[\max_{t}\bigl[I_{wtr}\left(\mathtt{x}_{ap},\mathtt{x}_{sta},t\right)\bigr]\right]_{\mathtt{x}_{ap},\mathtt{x}_{sta}}\nonumber\\
&= \lambda_{ap} \lambda_{sta} P_T \displaystyle \int_{\mathtt{x}_{ap} \in \mathbf{R}^2_{reg}} \displaystyle \int_{\mathtt{x}_{sta} \in \mathbf{R}^2_{net}} l\left(\|\overrightarrow{\mathcal{O}Q}\|\right) G_T\left(\theta_w\right) G_R\left(\theta_r\left(t_0\right)\right) d\mathtt{x}_{sta} d\mathtt{x}_{ap}, {\rm{~~}} t_{n} \le t \le t_{n+1}
\end{align}

\noindent where $t_0 = \arg\max_{t} I_{wtr}\left(\mathtt{x}_{ap},\mathtt{x}_{sta},t\right), {\rm{~}} t_{n} \le t \le t_{n+1}$.

\subsubsection{Radar-to-Wi-Fi (RtW) interference}
A RtW interference is defined as an \textit{average} radar signal power received by all the Wi-Fi RXs from the radar within its one rotation time. Changing (\ref{eq_i_individual}) to indicate that the radar is the interfering TX and a Wi-Fi node is the victim RX, a RtW interference can be formulated as
\begin{align}\label{eq_i_average}
I_{rtw}\left(\mathtt{x}_{ap},\mathtt{x}_{sta},t\right) &= \frac{1}{\lambda_{ap}} \displaystyle \sum_{\mathtt{x}_{ap} \in \mathbf{R}^2_{reg}} {I \left(\mathtt{x}_{ap},\mathtt{x}_{sta},t\right)}\nonumber\\
&= \frac{P_T}{\lambda_{ap}} \displaystyle \sum_{\mathtt{x}_{ap} \in \mathbf{R}^2_{reg}} l\left(\|\overrightarrow{\mathcal{O}Q}\|\right) G_T\left(\theta_r\left(t\right)\right) G_R\left(\theta_w\right).
\end{align}
Now performance of a Wi-Fi RX is
\begin{equation}\label{eq_sinr_no_mitigation}
\mathtt{SINR}\left(\mathtt{x}_{ap},\mathtt{x}_{sta},t\right) = \frac{P_T G_T G_R l_w\left(\|\overrightarrow{\mathtt{x}_{ap}\mathtt{x}_{sta}}\|\right)}{I_{rtw}\left(\mathtt{x}_{ap},\mathtt{x}_{sta},t\right) + N_0}
\end{equation}
where $l_w\left(\cdot\right)$ is a 3rd Generation Partnership Project (3GPP) Urban Micro (UMi) path loss model for a small cell environment \cite{tr36814}. Note that this $\mathtt{SINR}$ is used to finally obtain (\ref{eq_nppi}).

\begin{figure}[t]
\centering
\includegraphics[width = 0.8\linewidth]{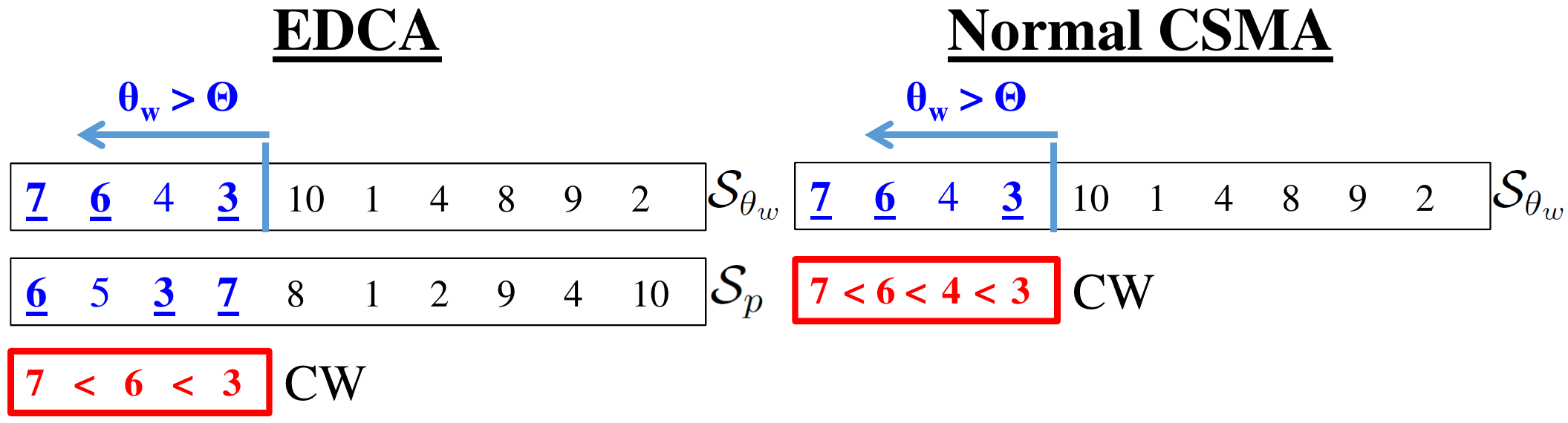}
\caption{Example of the proposed mechanism ($\lambda_{sta} = 10$)}
\label{fig_sets}
\end{figure}

\section{WtR Interference Mitigation}\label{sec_mitigation}
In this section, we propose a WtR interference mitigation method. Distinguished from \cite{zander_dyspan14}-\cite{hanbat16}, it enables every Wi-Fi network to keep operation, with the nodes with its beams sufficiently \textit{off from the interference axis}; that is, $\theta_w > \Theta$ where $\Theta$ is the threshold of $\theta_w$. The unnecessity of ceasing operation by any Wi-Fi network is the key benefit that this method introduces and thus keeps the Wi-Fi performance at an operable level.

\subsection{Radar Sweep Period, $T_{sweep}$}
A radar beam faces a Wi-Fi region for only a proportion of time within a rotation. We divide a rotation time into ``sweep'' and ``safe'' periods, denoted by $T_{sweep}$ and $T_{safe}$, respectively. During a $T_{safe}$, no interference mitigation is needed and thus the Wi-Fi TXs can access the medium as described in Section \ref{sec_system_model}. On the other hand, for a $T_{sweep}$, only the Wi-Fi TXs with $\theta_w > \Theta$ are eligible for competition for the medium.

A $T_{sweep}$ is measured and periodically broadcasted to the Wi-Fi APs by the spectrum access system (SAS), the database mediating the radar and the communications system \cite{ro_35g}. In turn, an AP updates $T_{sweep}$ and broadcasts to the network using a beacon. For accurate measurement, a SAS sensor is almost co-located at the Wi-Fi networks so that a $T_{sweep}$ is evaluated approximately the same for the Wi-Fi nodes. At every transition from a $T_{safe}$ to $T_{sweep}$, there is a mitigation time, $\tau$. If a packet to be transmitted is shorter than $\tau$, the node is \textit{eligible} to participate in a competition for the medium; otherwise, the node must add $T_{sweep}$ to its backoff time.

\subsection{Wi-Fi Off-axis Angle, $\theta_{w}$}
Each Wi-Fi node (AP or STA) is able to compute $\theta_w$ autonomously, based on (i) \textit{location of the radar} provided via the SAS and (ii) its own position measured on its own. Whereas the location measurement method is beyond the scope of this letter, the assumption remains reasonable based on prior methods such as \cite{localization_infocom17}.

However, it is necessary to analyze the impacts of the inaccuracy on the radar performance. For instance, an inaccurate localization can lead a Wi-Fi node with a smaller $\theta_w$ to be chosen for transmission. It results in a higher WtR interference occurs than it should be, which consequently incurs a lower radar performance. Also, the Wi-Fi performance can be affected when the opposite is the case. A Wi-Fi TX with a large enough $\theta_{w}$ and a high priority can be excluded from transmission due to an inaccurate localization. This will incur performance degradation in Wi-Fi.

\subsection{Wi-Fi Protocol for WtR Interference Mitigation}
Assuming accurate localization of Wi-Fi nodes, we propose a hybrid coordination function (HCF) where a distributed coordination function (DCF) during a $T_{safe}$ and a point coordination function (PCF) during a $T_{sweep}$. For the EDCA, the eligible nodes are selected in the following manner. $\mathcal{S}_p$ and $\mathcal{S}_{\theta_w}$ are sets of indexes of the nodes in a network that are sorted in descending order in terms of priority and off-axis angle $\theta_w$, respectively, where $\mathbb{N}\left[\mathcal{S}_p\right] = \mathbb{N}\left[\mathcal{S}_{\theta_w}\right] = \lambda_{sta}$. Suppose that the first $m$ nodes of $\mathcal{S}_{\theta_w}$ meet the criterion $\theta_w > \Theta$. Then take the first $m$ nodes from $\mathcal{S}_{p}$ as well, and obtain the node indexes that belong to the $m$-element subsets of both sets. Fig. \ref{fig_sets} describes an example of this mechanism. Applying the off-axis angle criterion $\theta_w > \Theta$, nodes $\{6,5,3,7\}$ remain in $\mathcal{S}_p$ and $\{7,6,4,3\}$ remain in $\mathcal{S}_{\theta_w}$. Prioritizing $\mathcal{S}_{\theta_w}$ over $\mathcal{S}_{p}$, the intersection is sorted as $7 < 6 < 3$ and allocated smaller values of CW in that order. Note from Fig. \ref{fig_sets} that \textit{as $\Theta$ increases, $\mathbb{N}\left[\mathcal{S}_{\theta_w}\right]$ decreases and thus the intersection gets smaller as an immediate consequence}. For the normal CSMA, since no $\mathcal{S}_p$ is defined, nodes $\{7,6,4,3\}$ are eligible and CW values are allocated in the order of $7 < 6 < 4 < 3$.

A STA exploits a channel sounding event to report its location to the AP. Although channel sounding sacrifices throughput due to overhead, this protocol adopts an obligatory reporting policy as in the Dynamic Frequency Selection (DFS) of IEEE 802.11h, considering the significance of the radar operation. Our protocol suggests that every STA reports its location at least once within a radar ``revolution'' time.

\begin{table}[t]
\caption{Parameters}
\centering
\begin{tabular}{| l || c || c |}
\hline
\multicolumn{1}{|c||}{\textbf{Parameter}} & \multicolumn{1}{c||}{\textbf{Wi-Fi}} & \multicolumn{1}{c|}{\textbf{Radar}} \\ \hline \hline
Carrier frequency & \multicolumn{2}{c|}{3.5 GHz}\\ \hline
Bandwidth & 20 MHz & 10 MHz \\
TX power & 30 dBm (AP), 10 dBm (STA) & 90 dBm\\
Max antenna gain & 2.15 dBi per element, 4$\times$4 array ($\lambda/2$ array) & 33.5 dBi\\
Noise power (dBm) & -100.99 dBm & -104 dBm\\ \hline
\end{tabular}
\label{table_parameters}
\end{table}

\section{Numerical Results}\label{sec_numerical}
In this study, we distribute Wi-Fi networks in a region with area of $\left|\mathcal{R}_{reg}^2\right| \approx 3.14 \text{ km}^2$. The area of each Wi-Fi network is $\left|\mathcal{R}_{net}^2\right| \approx 0.04 \text{ km}^2$. We run 10,000 ``drops'' in MATLAB with the parameters that are summarized in Table \ref{table_parameters}.

In Fig. \ref{fig_results_inr}, we show that the proposed method leads to reduction of separation distance. We use interference-to-noise ratio (INR) to examine the separation distance where $\mathtt{MMAI}$ in (\ref{eq_i_aggregate_mean}) is used for the ``interference.'' Normal CSMA yields lower INR as it incurs lower probability that an AP wins the medium, whereas EDCA yields higher WtR interference due to higher probability of AP transmission. With the mitigation technique, the interference gap between EDCA and normal CSMA decreases since the mitigation techniques forces a network to consider the off-axis angles before the priority.

In Fig. \ref{fig_results_nppi}, we show a cumulative distribution function (CDF) of $\mathtt{NPPI}$ derived in (\ref{eq_nppi}) according to type of access and whether the interference mitigation method is applied. The interference mitigation method yields at maximum 15-dB $\mathtt{NPPI}$ degradation mainly due to (i) higher RtW interference and (ii) less chance of priority-based TX selection. The higher RtW interference under the mitigation mode is due to the fact that by having a TX not facing the radar, a RX points its RX beam at the radar with a smaller $\theta_w$. The EDCA yields higher $\mathtt{NPPI}$ as it guarantees higher values of priority of the Wi-Fi TXs.

\begin{figure}[t]
\minipage{0.49\textwidth}
\includegraphics[width=\linewidth]{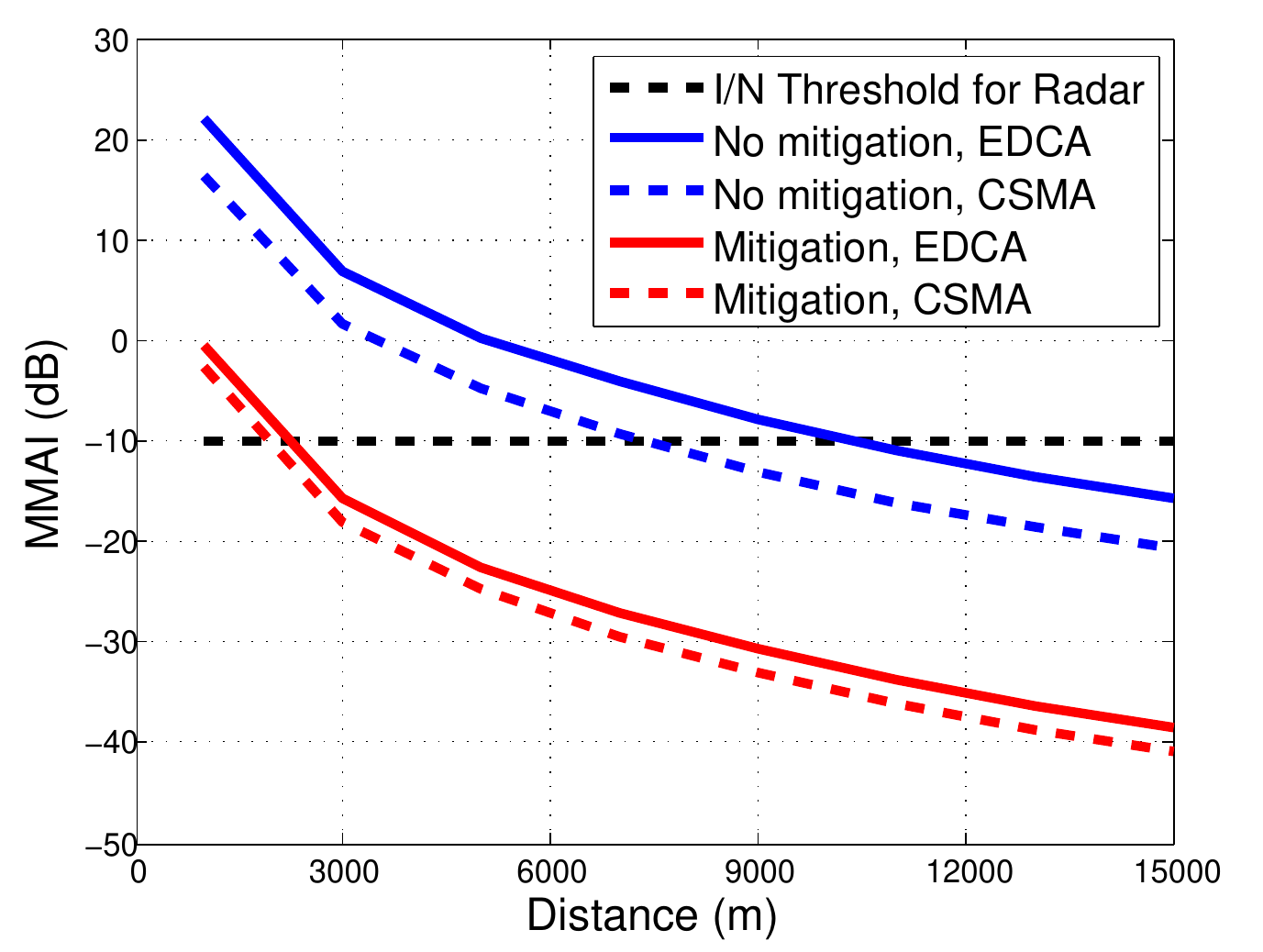}
\caption{$\mathtt{MMAI}$ vs. $d$ ($\lambda_{ap}=100, \lambda_{sta}=10$)}
 \label{fig_results_inr}
\endminipage\hfill
\minipage{0.49\textwidth}
\includegraphics[width=\linewidth]{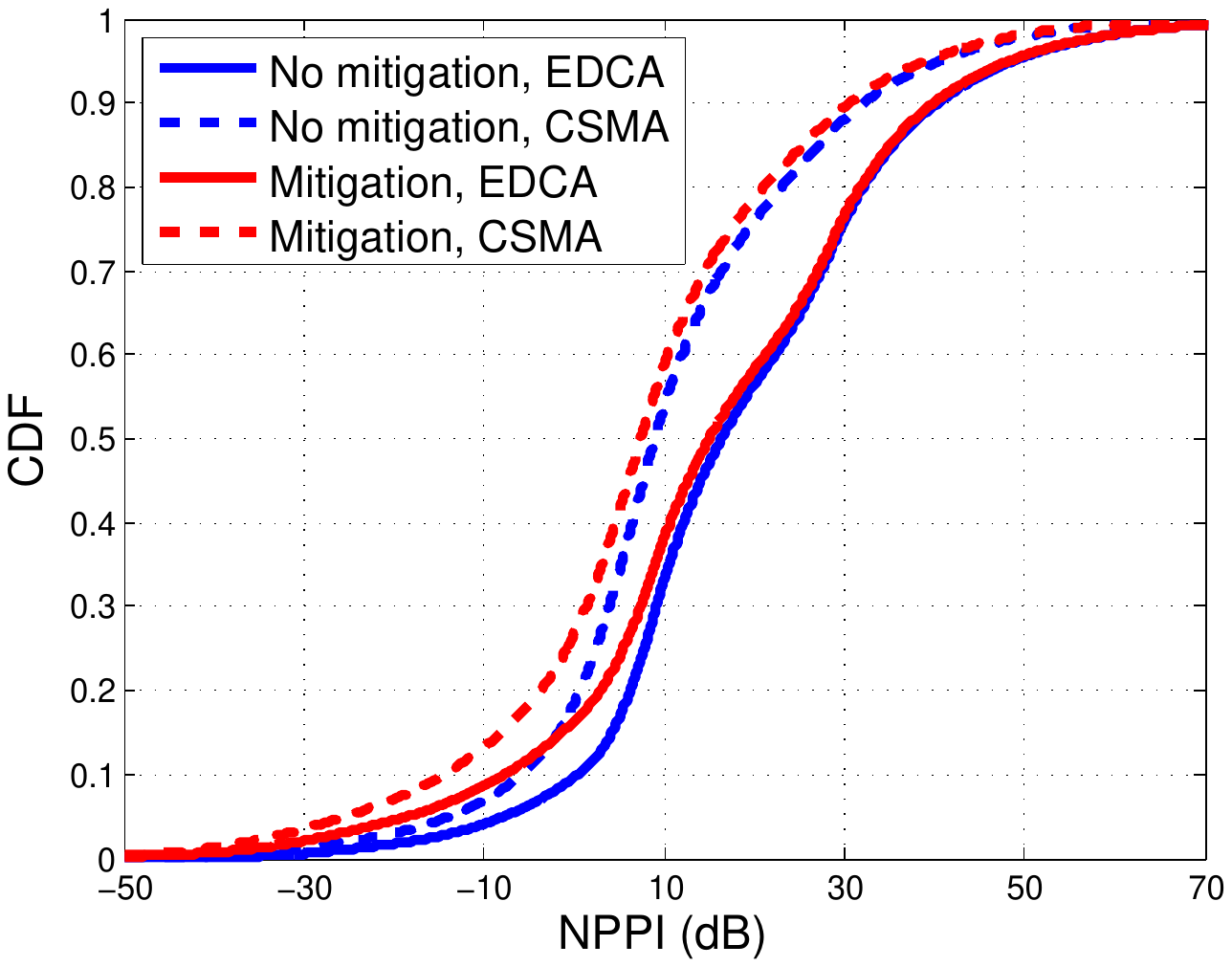}
\caption{CDF of $\mathtt{NPPI}$ ($\lambda_{ap}=100, \lambda_{sta}=10$)}
\label{fig_results_nppi}
\endminipage
\end{figure}

\begin{figure}[t]
\minipage{0.49\textwidth}
\includegraphics[width = \linewidth]{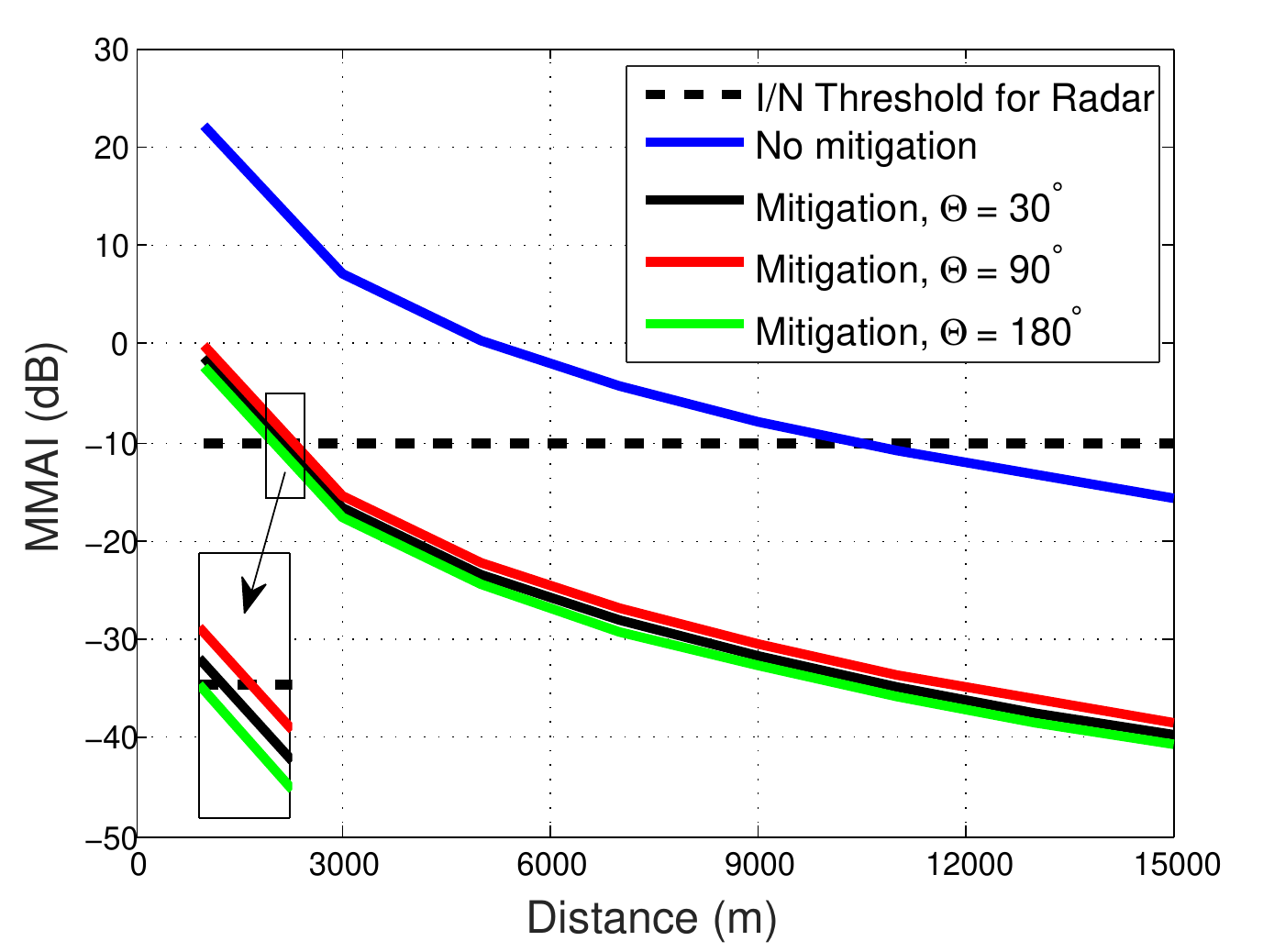}
\caption{$\mathtt{MMAI}$ vs. $\Theta$ ($\lambda_{ap}=100, \lambda_{sta}=10$, EDCA)}
\label{fig_results_inr_angle}
\endminipage\hfill
\minipage{0.49\textwidth}
\includegraphics[width = \linewidth]{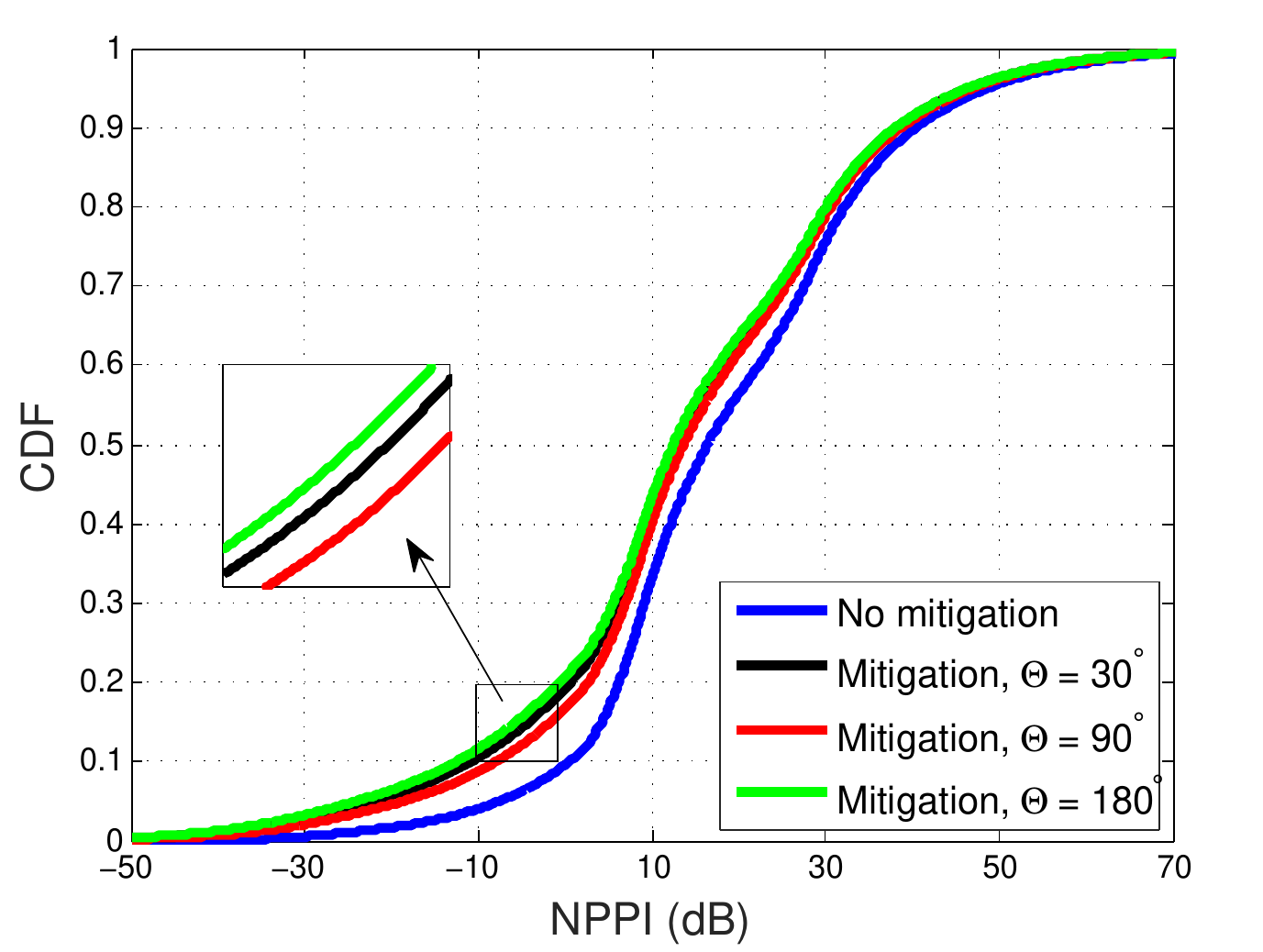}
\caption{CDF of $\mathtt{NPPI}$ for $\Theta$ ($\lambda_{ap}=100, \lambda_{sta}=10$, EDCA)}
\label{fig_results_nppi_angle}
\endminipage
\end{figure}

In Fig. \ref{fig_results_inr_angle}, we compare the WtR interference according to the off-axis angle threshold $\Theta$, with the EDCA. It is interesting that (i) variation of $\Theta$ has only little impact on WtR interference and (ii) the INR is in the order of $90^{\circ} > 30^{\circ} > 180^{\circ}$. The same tendency is shown in the $\mathtt{NPPI}$ that is given in Fig. \ref{fig_results_nppi_angle} as well. We discover that the WtR interference is a concave function according to $\Theta$, in which $\Theta = 90^{\circ}$ yields the maximum. As $\Theta = 0 \rightarrow 90^{\circ}$, the intersection between $\mathcal{S}_p$ and $\mathcal{S}_{\theta_w}$ becomes smaller as $\mathbb{N}\left[\mathcal{S}_{\theta_w}\right]$ set becomes smaller. Now the priority becomes the dominant criterion in selection of the TX. As a result, it is more probable that an AP becomes the TX. Therefore, (i) WtR interference increases due to higher interfering TX power, and (ii) $\mathtt{NPPI}$ increases by being more dominated by the priority. As $\Theta = 90 \rightarrow 180^{\circ}$, now it is very probable that no intersection exists between $\mathcal{S}_p$ and $\mathcal{S}_{\theta_w}$; hence, a TX is chosen in terms of $\theta_w$ only. As a result, (i) WtR interference decreases by having greater $\theta_w$'s, and (ii) $\mathtt{NPPI}$ decreases due to lower probability that an AP transmits.

\section{Conclusion}\label{sec_conclusion}
This letter proposes a technique that mitigates WtR interference while maintaining Wi-Fi system operation.


\end{document}